\documentclass[12pt]{article}
\usepackage{authblk}
\usepackage{amssymb}
\usepackage{amsmath}
\usepackage{graphicx}
\begin{document}
%\baselineskip 1 cm
%\baselineskip2.0 \baselineskip
\title{AdS braneworld with Backreaction}
\author[1,3]{Neven Bili\'c\thanks{bilic@irb.hr}}
\affil[1]{Rudjer Bo\v skovi\'c Institute, 10002 Zagreb, Croatia}
\author[2,4]{Gary B.\ Tupper\thanks{gary.tupper@uct.ac.za}}
\affil[2]{Centre  for Theoretical and Mathematical Physics, Department of physics, University of Cape Town,
Rondebosch~7701, South Africa}
\affil[3]{Departamento de F\'isica,
Universidade Federal de Juiz de Fora, 36036-330,
Juiz de Fora, MG, Brazil}
\affil[4]{Associate Member, National Institute for Theoretical Physics}

\maketitle

\begin{abstract}
We review the tachyon model derived from the dynamics of a 3-brane moving in the AdS$_5$ bulk.
The bulk geometry is based on the  Randall--Sundrum II model extended to include
the radion.
The effective tachyon Lagrangian is modified due to the back-reaction
of the brane on the bulk geometry.
\end{abstract}

%\maketitle

%\newpage

%\baselineskip 2.0 \baselineskip

%\section{Introduction}
\section{Introduction}

Branewarld cosmology is based on the  scenario in which matter is confined 
on a brane moving in the higher dimensional bulk
with only gravity allowed to propagate in the bulk  \cite{arkani,antoniadis,randall1,randall2}.
It is usually assumed that extra dimensions are compact and if their size 
is large enough compared to the Planck scale, such a scenario may explain 
the large mass hierarchy between the electroweak scale and the
fundamental scale of gravity. 
The Randall--Sundrum  solution \cite{randall1} to the hierarchy problem is a five dimensional 
universe containing two four dimensional branes with opposite brane tensions separated in the fifth dimension: 
the observer's brane is placed on the negative tension brane
and the separation 
  proposed is such that the strength of gravity on observer's brane 
is equal to the observed four-dimensional Newtonian gravity.
At the same time it was realized that the Randall--Sundrum model, as well as any similar braneworld model, 
may have interesting cosmological
implications \cite{flanagan}. In particular, owing to the presence of an extra dimension  
and the AdS$_5$ bulk cosmological constant related to the brane tension, the usual 
Friedmann equations are modified
\cite{binetruy}
so the model can have predictions different from the standard cosmology
and is therefore subject to cosmological tests \cite{godlowski}.

 In the second Randall--Sundrum model (RSII) \cite{randall2}
the negative tension brane is pushed off to infinity in the fifth dimension
and the Planck mass scale is determined by
the curvature of the five-dimensional space-time rather then the size of the fifth dimension.
Hence, the model  provides an alternative to compactification \cite{randall2}.
In RSII the bulk metric is AdS$_5/Z_2$ 
\begin{equation}
ds^2_{(5)}=e^{-2ky}g^{\mu\nu}dx^\mu dx^\nu -dy^2
 \label{eq3001}
\end{equation}
with the observer brane at $y=0$ and a negative tension brane at 
the AdS horizon at $y=\infty$.
The fifth dimension can be integrated out to obtain a purely four-dimensional action
with a well defined value for the Planck mass  of the order
$m_{\rm Pl}^2 \simeq (kK_{(5)})^{-1}$.

The new degree of freedom corresponding  to the fluctuations
 of the interbrane distance  along the extra dimension
implies the existence of a massless
scalar field:  the {\em radion} which may cause a distortion of 
the simple  AdS$_5$ geometry.
Besides, the correct description must also include 
matter on observer´s brane which also 
distorts the naive bulk geometry \cite{kim1,kim2}
(see also \cite{bagger}).

Various technical and phenomenological aspects of the radion have been 
extensively discussed.
Goldberger and Wise \cite{goldberger} proposed  a bulk scalar field propagating in the
background solution of the metric that generates a potential
that can stabilize the radion. The minimum of
the potential can be arranged to give the desired value of
the separation distance $d_5$ between the branes 
without fine-tuning of parameters. 
The mass and the wave function of the radion 
is determined including the back reaction of the bulk stabilization field on the metric \cite{csaki}, 
giving a typical radion mass of the order of the weak scale
 between 
0.100 and 1 TeV and the strength of its coupling
to the SM fields of the order of 1 TeV. 
Quite recently, it has been speculated that the evidence for the
 "Higgs boson" recently found at CERN may in fact  be 
the evidence for the radion \cite{cheung}.
 
In this paper  we investigate the dynamics of  a moving 3-brane 
in an extended second Randal Sundrum (RSII) model
which includes the back reaction due to the radion field.
A 3-brane moving in AdS$_5$ background of  the RSII model behaves effectively 
as a tachyon with the inverse quartic potential.
The RSII model may be  extended to include the back reaction due to the radion field. 
Then we show that the  tachyon Lagrangian is modified by the interaction with the radion 
and, as a consequence, the effective equation of state obtained 
by averaging over large scales describes a warm dark matter (DM).

\section{Gravity in the bulk}
\label{gravity}
Unless stated otherwise, we work in units $c=\hbar=1$ and keep the Newton constant $G$ 
explicit.
It is convenient to choose a coordinate system such that
$g_{(5)\mu 5}=0$ with metric 
%%EQUATION 1
\begin{eqnarray}
ds^2_{(5)} = g_{(5)MN}(X) dX^M dX^N 
=\Psi^2(x,y)g_{\mu\nu}(x) dx^\mu dx^\nu -\varphi^2
(x,y)dy^2 ,
\label{eq3002}
\end{eqnarray}
which admits Einstein spaces of constant
4-curvature.
Using (\ref{eq3002}) the bulk action may be expressed as \cite{kim1}
\begin{eqnarray}
S_{\rm bulk} & = & \frac{1}{K_{(5)}} \int d^5x \sqrt{g_{(5)}} \left[
-\frac{R_{(5)}}{2}-\Lambda_{(5)}\right] \nonumber\\%[0.5cm]
&=& \frac{1}{K_{(5)}} 
%\displaystyle{\int} d^4x \sqrt{-g} \displaystyle{\int} dy
\int d^4x \sqrt{-g}\int dy
\left[-\frac{R}{2}\Psi^2\varphi- 3g^{\mu\nu}
(\Psi\varphi)_{,\mu}\Psi_{,\nu} 
                + 6 \frac{\Psi^2(\partial_y\Psi)^2}{\varphi}
            -\Lambda_{(5)} \Psi^4\varphi \right] 
\label{eq3003}
\end{eqnarray}
The consistency with Einstein's equations outside the brane requires
%\vspace{-0.2cm}
\begin{equation}
R_{(5)\mu 5}  = 0.
\label{eq3005}
\end{equation}
This leads to 
\begin{equation}
\Psi = \exp \left(\int dy  \varphi \frac{\partial_y W}{W} \right) 
\label{eq3006}
\end{equation}
where the function  $W=W(y)$ is a background warp that does not depend on $x$.
A choice of  $\varphi$ 
(gauge choice)
is basically the  choice of parametrization of the distance along the
fifth dimension at fixed $x$.
It is convenient to impose the gauge condition
\begin{equation}
\Psi^2\varphi = W^2
\label{eq3007}
\end{equation}
so that the coefficient of $R$ in (\ref{eq3003})
is entirely fixed by the background.
With this gauge condition we find \cite{kim1}
\begin{equation}
\Psi(x,y) = \left[ W^{2} (y) + \phi(x) \right]^{1/2} , \hspace{1cm}
\varphi(x,y) = \frac{W^2(y)}{W^2(y)+\phi(x)}  .
\label{eq3008}
\end{equation}
where $\phi(x)$ is a function of $x$.
This yields
\begin{eqnarray}
S_{\rm bulk} & = & \frac{1}{K_{(5)}}\int d^4x \sqrt{-g}
\int dy \Biggl\{ -\frac{R}{2}W^2+\frac34 \frac{W^2}{(W^2+\phi)^2}
g^{\mu\nu}\phi_{,\mu}\phi_{,\nu} \Biggr. \nonumber\\ 
              &  & \Biggl. + \left [ 6(\partial_y W)^{2}- \Lambda_{(5)}W^{2} \right](W^2+\phi) \Biggr\}
\label{eq3009}
\end{eqnarray}
In order to keep a close connection with the Randall-Sundrum models, we take 
\begin{equation}
W=e^{-ky} \;\; {\rm on} \;\; 0\leq y \leq  l.
\end{equation}
The bulk metric is then given by
\begin{equation}
ds^2_{(5)}=(e^{-2ky}+\phi)g^{\mu\nu}dx^\mu  dx^\nu -\left(\frac{e^{-2ky}}{e^{-2ky}+\phi}\right)^2dy^2
 \label{eq2001}
\end{equation}
and the integration over $y$ yields
\footnote{Because the fifth dimension is S$^1/$Z the $y$-integrals  $\int_0^ldy$ are doubled.}
\begin{equation}
S_{\rm bulk} = \int d^4x \sqrt{-g}
\left\{ -\frac{R}{16\pi G}+\frac{3}{32\pi G} \omega(\phi)
g^{\mu\nu}\phi_{,\mu}\phi_{,\nu} 
 + \frac{3\tilde{k}}{K_{(5)}}\left[(1+\phi)^2-(e^{-2kl}+\phi)^2\right] \right\}
\label{eq3010}
\end{equation}
where we identified the four-dimensional Newton constant
\begin{equation}
 \frac{1}{8\pi G}= \frac{2}{K_{(5)}} \int_{0}^{l} dy W^{2} =
\frac{1 - e^{- 2 k l}}{k  K_{(5)}}.
\label{eq3011}
\end{equation}
Then the function $\omega$ is expressed as
\begin{equation}
\omega(\phi)= 16\pi G \int_{0}^{l} dy \frac{W^{2}}{(W^{2}+\phi)^2} =
\frac{1}{(1+\phi)(e^{-2kl}+\phi)},
\end{equation} 
and we use the abbreviation 
\begin{equation}
\tilde{k}= k-\frac{\Lambda_{(5)}}{6k} .
\end{equation}
The field $\phi(x)$ dubbed ``radion'' parameterizes the interbrane distance at fixed $x^\mu$ 
\begin{equation}
d_5=  \int_{0}^{l}dy\varphi = \int_{0}^{l} dy \frac{W^{2}}{W^{2}+\phi} =
\frac{1}{2k} \ln \frac{1+\phi}{e^{-2kl}+\phi} ,
\label{eq3013}
\end{equation} 
so that the distance to the AdS horizon $\displaystyle \lim_{l\rightarrow \infty} d_5$ 
remains finite. 
As in the RSII model, the metric \eqref{eq2001} will be a solution
to Einstein's equations provided
\begin{equation}
k^2 =-\frac{\Lambda_{(5)}}{6} ,
\end{equation}
where $\Lambda_{(5)}$ on the right-hand side is negative for AdS$_5$.

The bulk action (\ref{eq3010}) may be further simplified.
First, as we shall shortly see, the last term in curly brackets in (\ref{eq3010}) is canceled by
the brane action if the RSII fine tuning is imposed. Second, the radion kinetic term
 may be brought to the standard form  by introducing the canonically normalized radion
$\Phi$
via the transformation \cite{kim1}
\begin{equation}
\phi = (1 + e^{- 2 k l}) 
\sinh^{2}  \left( \sqrt{\frac{4\pi G}{3}} \Phi \right) +
e^{- k l}  \sinh  \left( \sqrt{ \frac{16\pi G}{3} }  \Phi \right) .
\label{eq3014}
\end{equation}
Then, the bulk action takes a simple form
\begin{equation}
S_{\rm bulk} = \int d^4x \sqrt{-g}
\left( -\frac{R}{16\pi G}+\frac12
g^{\mu\nu}\Phi_{,\mu}\Phi_{,\nu} 
\right)
 \label{eq3015}
\end{equation}

\section{Brane action}
Consider a 3-brane 
moving in the 4+1 bulk spacetime with metric (\ref{eq2001}).
The points on the brane are parameterized by
$X^{\mu} (x^{\mu})$, and $g^{\rm ind}_{\mu \nu} =
g_{(5) MN}  X_{, \mu}^{M}  X_{, \nu}^{N}$ is the induced metric.
Taking the Gaussian normal parameterization 
\begin{equation}
 X^M(x^\mu)=\left(x^\mu, y(x^\mu)\right)
\end{equation}
we have
\begin{equation}
g^{\rm ind}_{\mu \nu}=\left(\frac{e^{-2ky}}{e^{-2ky}+\phi}\right)^2
\left[\frac{(e^{-2ky}+\phi)^3}{(e^{-2ky})^2}g_{\mu\nu} -y_{,\mu} y_{,\nu} \right]
 \label{eq2002}
\end{equation}
The brane action is then given by
\begin{equation}
S_{\rm brane} =
-\sigma  \int d^{4}x \sqrt{- \det g^{\rm ind}_{\mu \nu}} 
  =- \sigma  \int  d^{4}x  \sqrt{-g} \,
(e^{-2ky}+\phi)^2 \left(
1-\frac{(e^{-2ky})^2}{(e^{-2ky}+\phi)^3}g^{\mu\nu}y_{,\mu} y_{,\nu}\right)^{1/2} 
 \label{eq2003}
\end{equation}
%\subsubsection{Fixed branes}
From this we find
the contribution of observer's brane  at $y=0$ and 
the negative tension brane at $y=l$ 
as 
\begin{equation}
S_{\rm brane}|_{y=0}
+ S_{\rm brane}|_{y=l} =
- \sigma_0  \int  d^{4}x  \sqrt{-g} \, (1+\phi)^2
- \sigma_l  \int  d^{4}x  \sqrt{-g} \,(e^{-2kl}+\phi)^2
 \label{eq2004}
\end{equation}
With the RSII fine tuning
\begin{equation}
\sigma_0=-\sigma = \frac{3\tilde{k}}{K_{(5)}}=\frac{6k}{K_{(5)}}
 \label{eq2005}
\end{equation}
the brane contributions cancel the last term on the right-hand side of
(\ref{eq3010}).

Hence, the appearance of a massless mode -– the radion -– causes 
two effects. First, according to (\ref{eq2002}), matter on observer's brane sees the (induced) metric 
\begin{equation}
\tilde{g}_{\mu\nu}=\left.g^{\rm ind}_{\mu \nu}\right|_{y=0}=
(1+\phi)g_{\mu\nu}
 \label{eq4002}
\end{equation} 
 and second,
the physical distance to the AdS$_5$ horizon at coordinate infinity
\begin{equation}
d_5=  \frac{1}{2k} \ln \frac{1+\phi}{\phi} ,
%\label{eq3013}
\end{equation} 
is no longer infinite if $\phi\neq 0$. 
The physical size of the 5-th dimension is of the order 
$1/k \sim  l_{\rm Pl}$ although its coordinate size is infinite. 

\subsection{Dynamical brane as a tachyon}
Consider an additional 3-brane moving in the bulk with 
metric (\ref{eq2001}). In this case, the fifth coordinate  is 
treated as a dynamical scalar  field $y(x)$.
Changing $y(x)$ to a new field 
\begin{equation}
\theta(x)=e^{ky(x)}/k
\label{eq1106}
\end{equation}
from (\ref{eq2003}) we obtain \cite{bilic3}
\begin{equation}
S_{\rm brane} =
 - \int  d^{4}x  \sqrt{-g} \,
\frac{\sigma}{k^4\theta^4}(1+k^2\theta^2 \phi)^2
\sqrt{1-\frac{g^{\mu\nu}\theta_{,\mu}\theta_{,\nu}}{(1+k^2\theta^2\phi)^3}}.
 \label{eq2006}
\end{equation}

When $\phi=0$ we have the pure undistorted AdS$_5$ and 
\begin{equation}
S_{\rm brane}^{(0)} =
 - \int  d^{4}x  \sqrt{-g} \,
\frac{\sigma}{k^4\theta^4}
\sqrt{1-g^{\mu\nu}\theta_{,\mu}\theta_{,\nu}}
 \label{eq2007}
\end{equation}
This action describes a tachyon with inverse quartic potential.
A related model is discussed by Silverstein and Tong \cite{silverstein}
where a D3-brane action is given by
\begin{equation}
S_{\rm D3} =
 \int  d^{4}x  \sqrt{-g} \,
\frac{\sigma}{k^4\theta^4}
\left[1-
\sqrt{1-g^{\mu\nu}\theta_{,\mu}\theta_{,\nu}}
\right]
 \label{eq2008}
\end{equation}
In this case, the pressure $p={\cal L}$ is positive definite
so there is no dark energy resulting
(at low ``velocity'' there is no force on the D-brane).
Although in our case (\ref{eq2007}) $p<0$,
the steep potential drives a dark matter attractor
\cite{abramo}
so $p\rightarrow 0^-$ very quickly and this 
``tachyon dust'' clusters efficiently on caustics
\cite{felder}.
One can get inflation or DE by adding a potential
term $V(\phi)$ to
(\ref{eq2008})
but that is somewhat ad-hoc.
Reversing the brane charge ($\overline{\rm D3}$-brane) 
in (\ref{eq2008}) 
gives $p<0$ but the steepness of the potential remains an obstruction.
What does make the tachyon intriguing is that even if $k^{-1} \sim l_{\rm Pl}$
as the AdS horizon is approached $e^{2ky}$ may be so large that
$\theta$ is ${\cal O}(H^{-1})$ without any fine tuning or dimensionful parameters.

On the other hand, if $\phi$ is not strictly zero
within ${\cal L}$,
the tachyon can drive a transition from
$k^2\theta^2\phi\ll 1$ regime to $k^2\theta^2\phi\gg 1$. 
In the latter regime 
the brane action (\ref{eq2006}) takes the form 
\begin{equation}
S_{\rm brane}\simeq
 - \int  d^{4}x  \sqrt{-g} \,
\sigma \phi^2
\sqrt{1-g^{\mu\nu}\bar{\theta}_{,\mu}\bar{\theta}_{,\nu}}
 \label{eq3107}
\end{equation}
where $\bar{\theta}_{,\mu}=\theta_{,\mu}/(k^3\theta^3\phi^{3/2})$.
One sees an obvious similarity to the Chaplygin gas \cite{kam9,bilic,fab19,ben32}:
The Hubble drag drives the brane velocity towards vanishing such that $\sigma\phi^2$
serves as a variable tension, or potential for $\theta$ through an implicit dependence of $\phi$
on $\theta$. The latter is similar to "quartessence" \cite{mak11}, the model for DE/DM unification.
Although not a single field model, this two component model has a potential to give
both DE and DM out of a single geometric structure. 

\subsection{Pressureless matter on the $y=0$ brane}
If matter is placed on the $y=0$ brane, its action is
\begin{equation}
S_{\rm matt}=
\int  d^{4}x  \sqrt{-\tilde{g}} \, {\cal L}_{\rm matt}
 \label{eq4001}
\end{equation} 
where $\tilde{g}$ is the determinant of the metric 
(\ref{eq4002})
 induced on the
$y=0$ brane. 
Pressureless matter can be modeled using a complex scalar field.
Consider a  Lagrangian of the type
\begin{equation}
{\cal{L}}_{\rm matt}
 =   g^{\mu \nu} {\Psi^*}_{, \mu}
\Psi_{, \nu}
 -V(m^2|\Psi|^{2})
\label{eq4003}
\end{equation}
for  a complex scalar field 
\begin{equation}
\Psi = \frac{\varphi}{\sqrt{2}\, m}
\exp (-i m \chi) ,
\end{equation}
 where $m$ is the mass appearing in the
potential $V$.
In the Thomas-Fermi approximation \cite{bilic}
the Lagrangian (\ref{eq4003}) becomes
\begin{equation}
{\cal{L}}_{\rm mattTF}
 = \frac{\varphi^2}{2} g^{\mu \nu}
 \chi_{,\mu} \chi_{,\nu}
 -V(\varphi^2/2) .
\label{eq4004}
\end{equation}
with the equations of motion for the fields
$\varphi$ and $\chi$
\begin{equation}
g^{\mu \nu} \chi_{,\mu} \chi_{,\nu} = V'(\varphi^2/2)\, ,
\label{eq4005}
\end{equation}
\begin{equation}
(\sqrt{-g}\,\varphi^2
 g^{\mu \nu} \chi_{,\nu} )_{,\mu}=0 ,
\label{eq4006}
\end{equation}
where $V'(x)=dV/dx$.
Assuming $V' >0$, the field $\chi$ may be treated as a
velocity potential for the fluid 4-velocity
\begin{equation}
u^{\mu} = g^{\mu \nu} \chi_{,\nu} / \sqrt{V'} \, ,
\label{eq4007}
\end{equation}
As a consequence,
 the stress-energy tensor $T^{\mu \nu}$ constructed
from the Lagrangian (\ref{eq4003})
takes the perfect fluid
form, with the parametric equation
of state
\begin{equation}
\rho = \frac{\varphi^{2}}{2} V'+ V ,
 \hspace{1cm} p = \frac{\varphi^{2}}{2}
V' - V  .
\label{eq4008}
\end{equation}
Now we assume $p=0$. In this case we obtain an equation
\begin{equation}
 \frac{\varphi^{2}}{2}
V' = V  .
\label{eq4009}
\end{equation}
with solution
\begin{equation}
 V=\frac12 m^2\varphi^2 .
\label{eq4010}
\end{equation}
Defining a new field $\alpha=m^2\varphi^2$, redefining $\chi \rightarrow m\chi$,
and replacing $g^{\mu\nu}\rightarrow \tilde{g}^{\mu\nu}=g^{\mu\nu} (1+\phi)^{-1}$
we finally obtain the Lagrangian for pressureless matter as
\begin{equation}
{\cal{L}}_{\rm matt}
 = \frac{\alpha}{2} \left[(1+\phi)^{-1} g^{\mu \nu}
 \chi_{,\mu} \chi_{,\nu}
 -1\right] 
\label{eq4012}
\end{equation}
and the matter action as
\begin{equation}
S_{\rm matt}=
\int  d^{4}x  \sqrt{-g} \, \frac{\alpha}{2} \left[(1+\phi) g^{\mu \nu}
 \chi_{,\mu} \chi_{,\nu}
 -(1+\phi)^2\right] .
 \label{eq4013}
\end{equation} 
The field $\alpha$ is not dynamical and, as we shall shortly see,
will be eliminated from the field equations.

\section{Backreaction}
\label{back}
%\subsection{BW2013 representation}
%\label{bw2003}
The total action as seen on observer's brane is 
\begin{equation}
S = S_{\rm bulk} + S_{\rm brane}+S_{\rm matt},
\label{eq0001}
\end{equation}
where $S_{\rm bulk}$, $S_{\rm brane}$, and $S_{\rm matt}$ are defined in
(\ref{eq3015}), (\ref{eq2006}), and (\ref{eq4013}), respectively.
For the moment we ignore the  pressureless matter on observer's brane
and  let $l \rightarrow \infty$.
Then, the relation (\ref{eq3014}) between $\phi$ and $\Phi$ becomes
\begin{equation}
\phi =  
\sinh^{2}  \left( \sqrt{4\pi G/3} \, \Phi \right) ,
\label{eq0002}
\end{equation}
and the Newton constant defined in (\ref{eq3011}) is simply related to the bulk gravitational
constant as
\begin{equation}
 8\pi G = 
k K_{(5)} .
\label{eq0102}
\end{equation}
From now on we work in units $8\pi G=1$.
%Instead of the $\theta$ field defined in (\ref{eq1106}),
It is convenient to replace $\theta$ with a new field 
\begin{equation}
\Theta(x)
= 3e^{-2ky(x)}=\frac{3}{k^2\theta(x)^2}
\label{eq2108}
\end{equation}
and introduce new constants
\begin{equation} 
\lambda =\sigma /(6k^{2} ), \hspace{1cm}  \ell =\sqrt{6} /k .
\label{eq24} 
\end{equation} 
Then the combined radion and brane Lagrangian becomes
\begin{equation}
{\cal{L}}=\frac{1}{2}X
-\frac{\lambda}{\ell^2}\psi^2\sqrt{1-\ell^2\frac{Y}{\psi^3}} 
\label{eq2009}
\end{equation}
where we have used the abbreviations
\begin{equation}
X=g^{\mu\nu}\Phi_{,\mu}\Phi_{,\nu} ,
\hspace{1cm}
Y=g^{\mu\nu}\Theta_{,\mu}\Theta_{,\nu} ,
\label{eq2010}
\end{equation}
and
\begin{equation}
\psi=
2\Theta +6 \sinh^2  \left( \sqrt{\frac{1}{6}} \, \Phi \right) ,
\label{eq2012}
\end{equation}
The  energy-momentum tensor corresponding to the above Lagrangian
\begin{equation}
T_{\mu\nu} =2\frac{\delta{\cal L}}{\delta g^{\mu\nu}}-{\cal L}g_{\mu\nu}= \Phi_{,\mu} \Phi_{,\nu}+
\frac{\lambda}{\ell^2\psi}\frac{1}{\sqrt{1-\ell^2 Y/\psi^3}}\,\Theta_{,\mu}\Theta_{,\nu}
-{\cal{L}}g_{\mu\nu} ,
\label{eq2013}
\end{equation}
 may be expressed
as a sum of two components
\begin{equation}
T_{\mu\nu} = T_{1\mu\nu} 
+T_{2\mu\nu} 
%\label{eq0120}
\end{equation}
each representing a perfect fluid with
\begin{equation}
 T_{i\mu\nu}=
 (p_i+\rho_i)u_{i\mu} u_{i\nu} 
 -p_i g_{\mu\nu}, \hspace{1cm} i=1,2 .
%\label{eq0220}
\end{equation}
The corresponding velocities, pressures and densities are given by
\begin{equation}
u_{1\mu}=  \frac{\Phi_{,\mu}}{\sqrt{X}} ,
\hspace{1cm}
u_{2\mu}= \frac{\theta_{,\mu}}{\sqrt{Y}} ,
\label{eq5021}
\end{equation}
\begin{equation}
p_1=\frac{1}{2}X ,
\hspace{1cm}
p_2=-\frac{\lambda\psi^2}{\ell^2}\sqrt{1-\ell^2Y/\psi^3} ,
\label{eq3022}
\end{equation}
\begin{equation}
\rho_1=\frac{1}{2}X ,\hspace{1cm}
\rho_2=\frac{\lambda\psi^2}{\ell^2}\frac{1}{\sqrt{1-\ell^2Y/\psi^3}},
\label{eq4022}
\end{equation}

\subsection{Conjugate fields}
$\cal L$ and $T_{\mu\nu}$ may be expressed in terms of the conjugate 
fields (or conjugate ``momenta")
  $\pi_\Phi^\mu$ and  $\pi_\Theta^\mu$   defined as
\begin{equation}
\pi_\Phi^\mu=
\frac{\partial{\cal{L}}}{\partial\Phi_{,\mu}}=g^{\mu\nu}\Phi_{,\nu} ,
\label{eq2115}
\end{equation}
\begin{equation}
\pi_\Theta^\mu=
\frac{\partial{\cal{L}}}{\partial\Theta_{,\mu}}=
\frac{\lambda}{\psi}\frac{g^{\mu\nu}\Theta_{,\nu}}{\sqrt{1-\ell^2Y/\psi^3}} .
%\label{xxx}
\end{equation}
For timelike  $\Phi_{,\mu}$    and  $\Theta_{,\mu}$     we may also define the norms
\begin{equation}
\pi_\Phi=\sqrt{g_{\mu\nu}\pi_\Phi^\mu\pi_\Phi^\nu} , 
\hspace{1cm}
\pi_\Theta=\sqrt{g_{\mu\nu}\pi_\Theta^\mu\pi_\Theta^\nu}.
\label{eq2118}
\end{equation}
Using these equations one finds a useful expression 
\begin{equation}
1-\ell^2\frac{Y}{\psi^3}=
\frac{1}{1+\ell^2\pi_\Theta^2/(\lambda^2\psi)} .
\label{eq5008}
\end{equation}
Using (\ref{eq2115})-(\ref{eq5008}) we obtain
\begin{equation}
{\cal{L}}=\frac{1}{2}\pi_\Phi^2
-\frac{\lambda\psi^2}{\ell^2}\frac{1}{\sqrt{1+\ell^2\pi_\Theta^2/(\lambda^2\psi)}} ,
\label{eq2119}
\end{equation}
\begin{equation}
T_{\mu\nu} = \pi_{\Phi\mu} \pi_{\Phi\nu}+
\frac{\ell^2\psi}{\lambda}\frac{\pi_{\theta\mu} \pi_{\theta\nu}}{\sqrt{1+\ell^2\pi_\Theta^2/(\lambda\psi)}}
-g_{\mu\nu}{\cal{L}} .
\label{eq2120}
\end{equation}
and 
\begin{equation}
p_1=\frac{1}{2}\pi_\Phi^2;\hspace{0.7cm}
p_2=-\frac{\lambda\psi^2}{\ell^2}\frac{1}{\sqrt{1+\ell^2\pi_\Theta^2/(\lambda^2\psi)}};
\label{eq2122}
\end{equation}
\begin{equation}
\rho_1=\frac{1}{2}\pi_\Phi^2;\hspace{0.7cm}
\rho_2=\frac{\lambda\psi^2}{\ell^2}\sqrt{1+\ell^2\pi_\Theta^2/(\lambda^2\psi)}     ;
\label{eq2123}
\end{equation}
The same expression for $T_{\mu\nu}$ is obtained by making use of the canonical
definition
\begin{equation}
T_{\mu\nu}^{\rm can}=
\sum_{\varphi,\pi} \varphi_{,\mu}\pi_{\phi\nu} 
-{\cal{L}}g_{\mu\nu}
 \label{eq2107}
\end{equation} 
\subsection{Hamilton's equations}
The Hamiltonian may be identified with 
the total energy density 
\begin{equation}
{\cal{H}}=T^\mu_\mu +3{\cal{L}}=\rho_1+\rho_2,
 \label{eq2109}
\end{equation}
which yields
\begin{equation}
{\cal{H}}=\frac12\pi_\Phi^2
+\frac{\lambda\psi^2}{\ell^2}\sqrt{1+\ell^2\pi_\Theta^2/(\lambda^2\psi)} 
\label{eq2114}
\end{equation}
The Hamiltonian ${\cal H}$ (defined in \ref{eq2114} as a function of $\pi_\Phi^\mu$, $\pi^\mu_\Theta$,
$\Phi$, and $\Theta$)  is related to ${\cal L}$ (defined in (\ref{eq2009}) as a function of
$\Phi_{,\mu}$, $\Theta_{,\mu}$,  $\Phi$, $\Theta$) 
 through the Legendre transformation
\begin{equation}
{\cal H} (\pi^\mu, \varphi)= \sum_{\{\pi,\varphi\}} \pi^\mu\varphi_{,\mu} -{\cal L} (\varphi_{,\mu}, \varphi) ,
\label{eq2024}
\end{equation}
where
\begin{equation}
\varphi_{,\mu} =  \frac{\partial{\cal H}}{\partial\pi^{\mu}},
\label{eq2025}
\end{equation}
\begin{equation}
\pi^\mu =  \frac{\partial{\cal L}}{\partial\varphi_{,\mu}}.
\label{eq2026}
\end{equation}
In (\ref{eq2024})-(\ref{eq2026}) $\pi$ stands for $\pi_\Phi$ or $\pi_\Theta$,
and $\varphi$ stands for $\Phi$ or $\Theta$.
The first pair of Hamilton's equations is obtained by multiplying (\ref{eq2025}) by
$u_1^\mu$ and $u_2^\mu$ for $\Phi$ and $\Theta$ fields , respectively.
From (\ref{eq2114}) we derive 
\begin{equation}
u_1^\mu\Phi_{,\mu} \equiv  \dot{\Phi} = 
\frac{\partial{\cal H}}{\partial\pi_\Phi}
\label{eq2127}
\end{equation}
\begin{equation}
 u_2^\mu\theta_{,\mu} \equiv  \dot{\Theta} = 
\frac{\partial{\cal H}}{\partial\pi_\Theta}
\label{eq2128}
\end{equation}
The remaining two Hamilton's equations are obtained by applying the covariant divergence to
(\ref{eq2026}) and using the Euler-Lagrange equations 
\begin{equation}
\frac{\partial{\cal{L}}}{\partial\varphi}=
\left(\frac{\partial{\cal{L}}}{\partial\varphi_{,\mu}}\right)_{;\mu}.
\label{eq2029}
\end{equation}
Then, with the help of (\ref{eq5021}) we find
\begin{equation}
\dot{\pi}_\Phi+3H_1\pi_\Phi =
-\frac{\partial{\cal{H}}}{\partial\Phi} 
\label{eq2131}
\end{equation}
\begin{equation}
\dot{\pi}_\Theta+3H_2\pi_\Theta =
-\frac{\partial{\cal{H}}}{\partial\Theta}
\label{eq2132}
\end{equation}
The quantities  $H_i$, $ i=1,2$, are related to the expansions of $u_i$ 
\begin{equation}
3H_i={u_i^\mu}_{;\mu} . 
\label{eq3033}
\end{equation}
The set of equations (\ref{eq2127}), (\ref{eq2128}), (\ref{eq2131}), and (\ref{eq2132}) 
are solved assuming spatially flat FRW spacetime, in which case
\begin{equation}
H_1=H_2=H
\label{eq4031}
\end{equation}
where $H$ is the Hubble expansion rate. 

For a more complete description we add to the total 
Lagrangian the contribution of pressureless matter on the observer's brane 
(\ref{eq4012}). In this case, there is an additional contribution to the Hamiltonian 
\begin{equation}
{\cal{H}}_\chi =\frac{\pi_\chi^2}{2\alpha (1+\phi)} +\frac{\alpha}{2}(1+\phi)^2.
 \label{eq3113}
\end{equation}
where 
\begin{equation}
\pi_\chi=\sqrt{g_{\mu\nu} \pi_\chi^\mu \pi_\chi^\nu}  .
\label{eq1019}
\end{equation}
and $\pi_\chi^\mu$ is the conjugate momentum of the field $\chi$.
The non-dynamical field $\alpha$ can be eliminated by the Hamilton's equation
\begin{equation}
\frac{\partial{\cal{H}}}{\partial\alpha}=0 ,
%\label{eq0113}
\end{equation}
which follows from the Euler-Lagrange equation $\partial{\cal{L}}/\partial\alpha=0$ and (\ref{eq2024}).
Then we find the total Hamiltonian
\begin{equation}
{\cal{H}}=\frac12\pi_\Phi^2
+\frac{\lambda\psi^2}{\ell^2}\sqrt{1+\ell^2\pi_\Theta^2/(\lambda^2\psi)} 
+\pi_\chi\sqrt{1+\phi},
 \label{eq3114}
\end{equation}
and we have two additional Hamilton's equations
\begin{equation}
  \dot{\chi} 
=\frac{\partial{\cal H}}{\partial\pi_\chi},
\label{eq1028}
\end{equation}
\begin{equation}
\dot{\pi}_\chi+3H\pi_\chi = 0.
\label{eq1032}
\end{equation}

Finally,  from (\ref{eq2127})--(\ref{eq2132}) and (\ref{eq1028})--(\ref{eq1032})
using (\ref{eq3114})
we obtain the following set of equations
\begin{equation}
 \dot{\Phi} = 
\pi_\Phi
\label{eq3127}
\end{equation}
\begin{equation}
 \dot{\Theta} = 
\frac{\psi}{\lambda}\frac{\pi_\Theta}{\sqrt{1+\ell^2\pi_\Theta^2/(\lambda^2\psi)}}
\label{eq3128}
\end{equation}
\begin{equation}
 \dot{\chi} 
= \sqrt{1+\phi}
\label{eq2037}
\end{equation}
\begin{equation}
\dot{\pi}_\Phi=-3H\pi_\Phi 
-\frac{3}{\ell^2 \lambda}
\frac{4\lambda^2\psi+3\ell^2\pi_\Theta^2}{\sqrt{1+\ell^2\pi_\Theta^2/(\lambda^2\psi)}}\phi' 
-\pi_\chi\sqrt{\frac{1}{6} \phi}
\label{eq3131}
\end{equation}
\begin{equation}
\dot{\pi}_\Theta =-3H\pi_\Theta
-\frac{1}{\ell^2 \lambda}
\frac{4\lambda^2\psi+3\ell^2\pi_\Theta^2}{\sqrt{1+\ell^2\pi_\Theta^2/(\lambda^2\psi)}}
\label{eq3132}
\end{equation}
\begin{equation}
\dot{\pi}_\chi=-3H\pi_\chi ,
\label{eq2032}
\end{equation}
together with the Friedmann equation for the scale $a(t)$
\begin{equation} 
\frac{\dot{a}}{a}=H=\sqrt{\frac{1}{3} {\cal{H}}}
\label{eq3032}
\end{equation}
where $\phi$ is defined in   (\ref{eq0002}) and
\begin{equation}
\phi'=\sqrt{\frac{1}{6}}
\sinh \left (\sqrt{\frac{2}{3}} \Phi\right).
\label{eq3031}
\end{equation}
Equation (\ref{eq2032}) is easily solved for $a$
\begin{equation}
\pi_\chi= \frac{\pi_{\chi 0}}{a^3} ,
\label{eq2033}
\end{equation}
where $\pi_{\chi 0}$ is a constant which could be fixed by physics.
For example, we may require that the fraction of dust (which represents baryons)
today is about $0.05 \rho_{\rm cr}$. More precisely, at $t=t_0$ when $a(t_0)=1$
we require
\begin{equation}
\rho_\chi(t_0)\equiv \pi_{\chi 0}\sqrt{1+\phi (t_0)}=0.05 \frac{3}{8\pi G} H_0^2 .
\label{eq2034}
\end{equation}
\begin{figure}[ht]
%calculated by Backreaction Model.nb
\begin{center}
\includegraphics[width=0.7\textwidth,trim= 0 0cm 0 0cm]{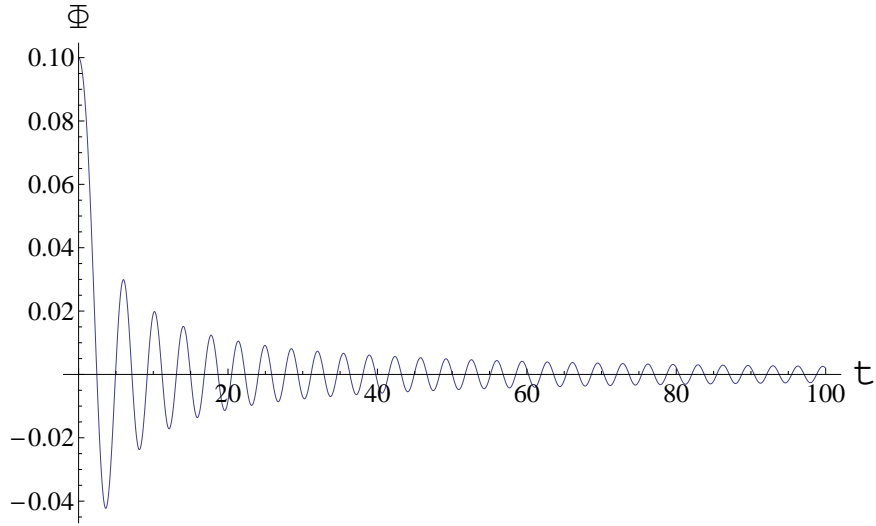}
%calculated by Backreaction Model.nb
\caption{Radion field in units of $\ell^{-1}$ as a function of time in  the backreaction model} 
 \label{fig1}
\end{center}
%\newpage
\end{figure}
\begin{figure}[ht]
\begin{center}
\includegraphics[width=0.7\textwidth,trim= 0 0cm 0 0cm]{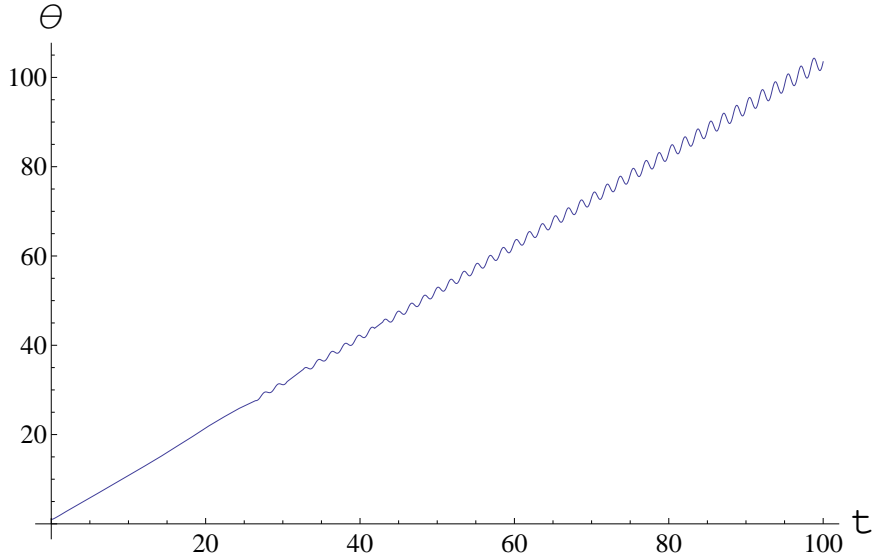}
%calculated by Backreaction Model.nb
\caption{Tachyon field $\theta=(2\Theta)^{-1/2}$ in units of $\ell$ as a function of time}
 \label{fig2}
\end{center}
\end{figure}
\begin{figure}[ht]
\begin{center}
\includegraphics[width=0.7\textwidth,trim= 0 0cm 0 0cm]{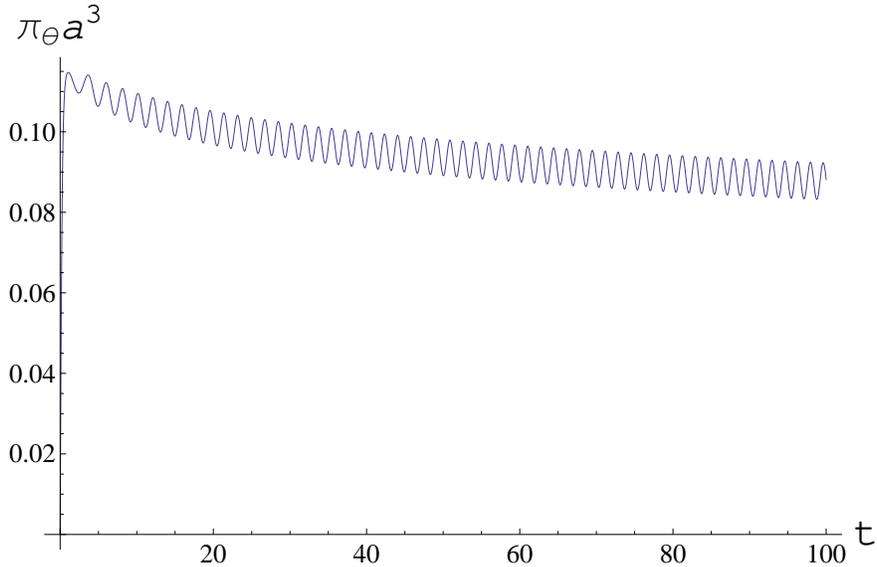}
%calculated by Backreaction Model.nb
\caption{Normalized conjugate momentum  (in units of $l^{-4}$)
associated with the field $\theta$   as a function of time} 
 \label{fig3}
\end{center}
\end{figure}
%%%%%%%%%%%%%%%%%%%%%%%%%%%%%%%%%%%%%%%%%
\begin{figure}[ht]
\begin{center}
\includegraphics[width=0.7\textwidth,trim= 0 0cm 0 0cm]{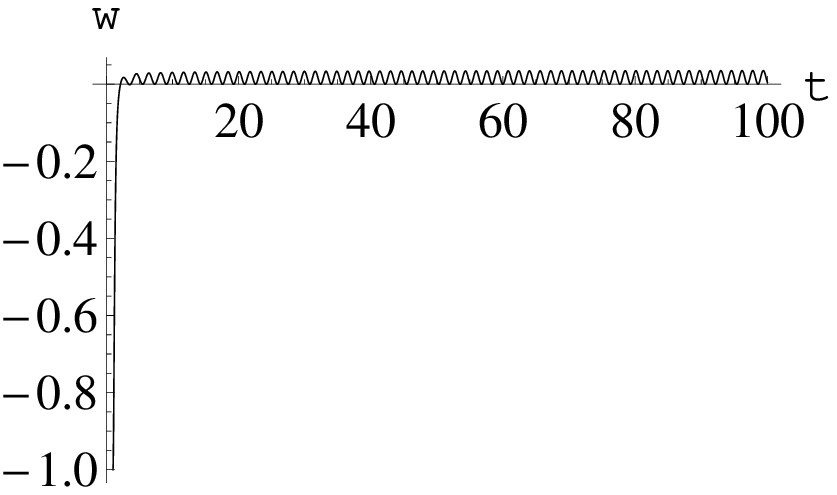}
%calculated by Warm Tachyon Numerics V4.nb
\caption{Evolution of the equation of state $w=p/\rho$ in the backreaction model} 
 \label{fig4}
\end{center}
\end{figure}
\section{Numerical results}
\label{numerical}

To exhibit the main features we neglect the dust on observer's brane and solve 
our equations assuming spatially flat FRW spacetime with line element  
\begin{equation}
ds^2=dt^2+a(t)^2(dr^2+r^2d\Omega^2)
\label{eq3034}
\end{equation}
We evolve the radion-tachyon system  in time measured 
in units of $\ell $ and we take  $\lambda \ell^2=1/3$. 
Equations (\ref{eq3127})-(\ref{eq3132}) are integrated 
starting from $t=0$ with the initial conditions $\Theta =1.01$, 
$\phi =0.1$, $\pi_\Phi =\pi_\Theta =0.00001$.
The results of integration are depicted in 
Figs.\ \ref{fig1}, \ref{fig2}, and \ref{fig3}.
%\noindent 
As one would anticipate from (\ref{eq3127}) with (\ref{eq3131}),
the field $\Phi$ undergoes damped oscillations with 
the amplitude decreasing as $1/t$ (Fig.\ \ref{fig1}).
In the asymptotic region one finds  an approximate solution \cite{bilic3}
\begin{equation}
\Phi =\frac{A}{t} \cos \frac{2t}{\ell} 
\label{eq4035}
\end{equation}
where $A$ is the amplitude of the asymptotic oscillations.  
 Comparing (\ref{eq4035}) with the exact solution for  
$\Phi$ depicted in Fig.\ \ref{fig1}, 
 we find  $A=0.1518$. 

As a consequence, the original tachyon field $\theta$ exhibits 
oscillations about a linear function which corresponds
to a  tachyon solution without radion (Fig.\ \ref{fig2}).
Similar oscillations are seen in 
the  momentum field 
\begin{equation}
\pi_\theta =-\ell\pi_\Theta(2\Theta)^{-3/2} 
\label{eq4034}
\end{equation}
conjugate to $\theta$ (Fig.\ \ref{fig3}).
To exhibit the oscillating behavior more clearly we have plotted 
$\pi_\theta$ multiplied by $a^3$.

After the transient period the equation of state $w=p/\rho$ becomes positive and oscillatory
(Fig.~\ref{fig4}).
In the asymptotic regime $t\rightarrow \infty$ we find 
an approximate expression \cite{bilic3}
\begin{equation} 
w \simeq 
\frac{\dot{\phi }^{2} }{\dot{\phi }^{2} +2 \psi^{3/2}  |\pi_\Theta| /\ell}
\end{equation} 
which yields
\begin{equation} 
w \simeq 
\frac{3}{2} A^2 \sin^2 \frac{2t}{\ell} 
\label{eq4036} 
\end{equation} 
to leading order in the amplitude $A$.
Since the oscillations in $w$ are rapid on cosmological timescales, 
it is most useful to time average co-moving quantities. 
The effective equation of state is then
\begin{equation} 
\left\langle p\right\rangle =\left\langle w\right\rangle \left\langle \rho \right\rangle ,
\end{equation} 
where $\langle x\rangle$ denotes the time average of the quantity $x$.
By averaging (\ref{eq4036}) over long timescales we find
\begin{equation} 
\left\langle w\right\rangle \simeq 
\frac{3}{4} A^2 =0.017
\label{eq4037} 
\end{equation}
This estimate hints at the analysis of  Avelino et al \cite{avelino} who
have recently shown that cosmological data favor 
a dark matter equation of state $w_{\rm DM} \approx 0.01$
rather then a pressureless, or cold dark matter equation of state.

The nature of dark matter (DM) is still an open question.
In spite of the large-scale successes of cold DM  there is still some
 unresolved issues such as  
overproduction of small scale structure and  halos with a central cusp 
\cite{frenk}. 
These problems are somewhat alleviated by warm DM and  in particular 
by sterile neutrino warm DM 
\cite{primack,boyanovsky}. However, a recent analysis \cite{schneider} shows that  
a realistic warm DM scenario with $m_{\rm DM} \simeq 4$ keV in agreement with recent 
constraints from Lyman-$\alpha$ forest \cite{viel} is not able to alleviate the small scale crisis
of cold DM structure formation. 

It is easy to demonstrate that the equation of state (\ref{eq4037}) may be associated with warm DM.
We assume that our equation of state corresponds to that of DM  thermal relics 
of mass $m_{{\rm DM}}$ at the time of 
radiation-matter equality  $t_{{\rm eq}}$.  Furthermore, assuming that DM particles 
constitute a non-relativistic gas at $t \sim t_{{\rm eq}}$,
the corresponding equation of state is, to a good approximation, 
 given by
\begin{equation}
w_{{\rm DM}} =\frac{T}{m_{{\rm DM}}} ,
\end{equation}
where $T$ is the temperature of the gas. Taking  $T=T_{\rm eq} =7.4$ eV at    $t=t_{\rm eq}$   
and identifying $w_{\rm DM}|_{\rm eq} =<w>=0.017$,  we obtain $m_{\rm DM} \simeq 430$ eV. 
These DM particles become non relativistic at the time when $T=T_{{\rm NR}} \simeq m_{{\rm DM}} $ 
corresponding to the cosmological scale
\begin{equation}
a_{{\rm NR}} \simeq  \frac{T_{\rm eq}}{T_{{\rm NR}}}= <w> a_{{\rm eq}} 
\end{equation} 
We next show that  
the horizon mass at the time when the equivalent DM particles 
just become non-relativistic is of the order 
typically of  a small galaxy. 
The horizon mass before equality evolves as  \cite{coles}
\begin{equation}
M_{{\rm H}} \simeq M_{{\rm eq}} \left(\frac{a}{a_{{\rm eq}} } \right)^{3} ,  
\end{equation}
where $M_{{\rm eq}} \simeq 2\times 10^{15} M_{\odot } $  for a spatially flat universe. 
Thus, at $a=a_{{\rm NR}}$ we obtain $M_{{\rm H}} \simeq 10^{10} M_{\odot }$ , 
the mass scale   typical  of a small galaxy and therefore the DM may be qualified as warm.

We have restricted attention to a homogeneous isotropic evolution for simplicity.
A warm DM model in general has a non-vanishing sound speed and hence may face the problem of
the well-known Jeans instability. The perturbations of the scale
smaller than the sonic horizon will be prevented from growing.
In our case, one cannot interpret $\sqrt{\langle w\rangle}$ as the adiabatic speed of perturbations.
Note also that the quantity $\dot{p}/\dot{\rho }$ cannot be identified with the speed of sound squared 
$c_{s}^{2} $ because $\dot{p}/\dot{\rho }$ is, in our case,   not positive semi-definite owing to interactions. 
The non-interacting radion is stiff matter, with unit speed of sound, 
whereas the non-interacting tachyon asymptotically has vanishing speed of sound.
As we have shown in appendix \ref{adiabatic} the sound 
speed squared for the composite is the sum of the 
components weighted by their fraction of $\rho +p$:
%%%%%%%%%%%%%%%%%%%%%%%%%%%%%%%%%%%%%%%%
\begin{equation}
 c_s^2
 =c_{s1}^2\frac{X{\cal L}_X}{X{\cal L}_X+Y{\cal L}_Y} 
+c_{s2}^2\frac{Y{\cal L}_Y}{X{\cal L}_X+Y{\cal L}_Y} 
 \label{eq6602}
\end{equation}
where $c_{s1}$ and $c_{s2}$ are defined as
\begin{equation}
c_{s1}^2=\frac{{\cal L}_X}{{\cal L}_X+2X{\cal L}_{XX}} ;
\hspace{1cm}
c_{s2}^2=\frac{{\cal L}_Y}{{\cal L}_Y+2Y{\cal L}_{YY}} .
\label{eq014}
\end{equation}
The expression  (\ref{eq6602}) agrees  with the speed of sound for a 
multicomponent fluid defined in \cite{tsagas}.
 
For the Lagrangian (\ref{eq2009})
we have ${\cal L}_X=1/2$, ${\cal L}_{XX}=0$, and
\begin{equation}
{\cal L}_Y =\frac{\lambda}{2\psi}\frac{1}{\sqrt{1-\ell^2 Y/\psi^3}}
=\frac{\lambda}{2\psi}\sqrt{1+\ell^2\pi_\Theta^2/(\lambda^2\psi)} ,
\label{eq1044}
\end{equation} 
\begin{equation}
Y{\cal L}_{YY} =\frac{\lambda}{4\psi^4}\frac{\ell^2Y}{(1-\ell^2 Y/\psi^3)^{3/2}}
=\frac{\ell^2\pi_\Theta^2}{4\lambda\psi^2}\sqrt{1+\ell^2\pi_\Theta^2/(\lambda^2\psi)} .
\label{eq2044}
\end{equation} 
Using this we obtain
 \begin{equation}
c_s^2=1-
\frac{\ell^2\pi_\Theta^4}{(\pi_\Theta^2+\pi_\Phi^2\sqrt{\lambda^2/\psi^2+\ell^2\pi_\Theta^2/\psi^3})
(\ell^2\pi_\Theta^2+\lambda^2\psi)}
\label{eq0049}
\end{equation}

\begin{figure}[ht]
\begin{center}
\includegraphics[width=0.7\textwidth,trim= 0 0cm 0 0cm]{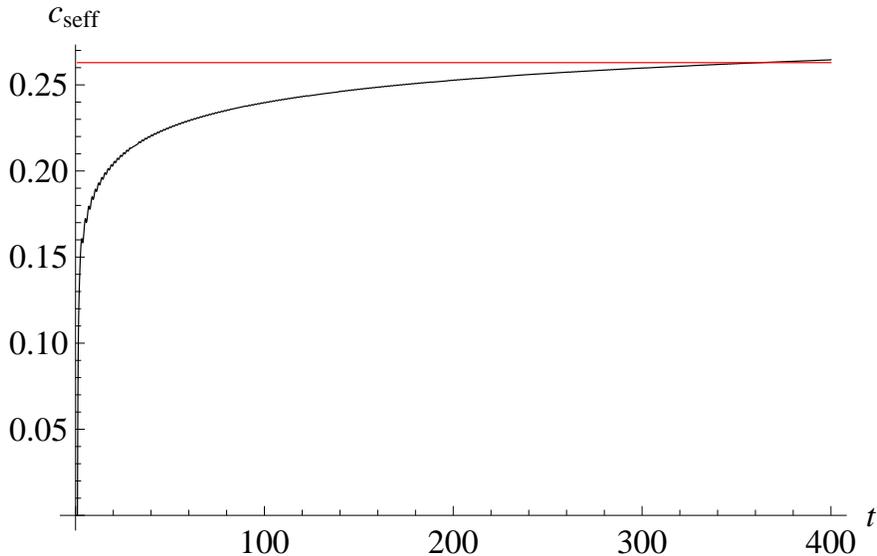}
\caption{Effective speed of sound. The horizontal red line represents the approximate 
asymptote given by (\ref{eq4044}).}
\end{center}
\label{fig5}
\end{figure}

 Due to the rapid oscillations, 
it is more appropriate to define the effective speed of sound as the ratio of the co-moving acoustic 
 to the co-moving particle horizon radii: 
\begin{equation}  
c_{s{\rm eff}} =\frac{\int dt c_{s}/a}{\int dt/a} .  
\label{eq4045}
\end{equation} 
 In Fig.~\ref{fig5} we plot the effective speed of sound defined in (\ref{eq4045})
together with the approximate asymptotic value 
\begin{equation} 
\left. c_{s{\rm eff}}\right|_{\rm app} \simeq \sqrt{3} A .
\label{eq4044} 
\end{equation} 
Note that  $\left. c_{s{\rm eff}}\right|_{\rm app}$ is twice as large as the ``average'' speed of sound
that one would naively expect from the  equation of state (\ref{eq4037}).

\section{Summary and Conclusions}

We have presented a derivation of the  effective tachyon Lagrangian 
 in an AdS$_5$ geometry distorted by the radion back-reaction.
The usual tachyon inverse quartic potential is modified due to the interaction with the radion.
The field equation of the resulting tachyon-–radion system is solved assuming homogeneous and isotropic evolution.
The back-reaction causes the tachyon-–radion system to behave as ``warm" tachyon matter with a linear barotropic equation of state. 

 In addition, we have studied the sound speed in our model.
We have derived the adiabatic speed of sound for a general model with two dynamical fields
and Lagrangian that depends on the two  kinetic terms and 
on the composite field in the form of a general function of the two fields.
We have shown that the effective sound speed in our model approaches 
asymptotically a constant value of the order of 0.25. 

The ultimate question regards the clustering properties of the model.  
A fluid with a nonzero sound speed has a characteristic scale below which
the pressure effectively opposes gravity. 
At the linear level one expects a suppression of small-scale structure formation: 
initially growing modes undergo damped oscillations once they enter the co-moving acoustic horizon. 
Perturbation theory is not the whole story -- it would be worth studying  
the nonlinear effects, e.g., using the Press-Schechter formalism as in 
the pure tachyon model of \cite{bilic?}.

\subsection*{Acknowledgments}

%\noindent 
This work was supported by the National Institute of Theoretical Physics of South Africa.
The work of G.B.T.\ was supported by a grant from the National Research Foundation of South Africa. 
The work of N.B.\ was partially supported by the Ministry of Science, 
Education and Sport of the Republic of Croatia under Contract No.\ 098-0982930-2864, and 
by the ICTP-SEENET-MTP grant PRJ-09 ``Strings and Cosmology`` 
in the frame of the SEENET-MTP Network.
N.B.\ thanks CNPq, Brazil, for partial support and the University of Juiz de Fora where a part of 
this work was completed.

\appendix

\numberwithin{equation}{section}

\section{Adiabatic speed of sound}
\label{adiabatic}
The derivation of the so called {\em adiabatic speed of sound} follows 
the procedure described in  Appendix A of \cite{bilic?} 
generalized to two dynamical fields $\Phi$ and $\Theta$.
Consider a Lagrangian of the form
\begin{equation}
{\cal{L}}={\cal{L}}(X,Y,\psi) ,
\label{eq1000}
\end{equation}
where $\psi=\psi(\Phi,\Theta)$ is an arbitrary function of  $\Phi$ and $\Theta$.
and 
\begin{equation}
X=g^{\mu\nu} \Phi_{,\mu} \Phi_{,\nu}, \hspace{1cm} Y=g^{\mu\nu} \Theta_{,\mu} \Theta_{,\nu},
\label{eq2000}
\end{equation}
For simplicity, we assume that the functional dependence of ${\cal{L}}$ on $X$ and $Y$ is such that
\begin{equation}
{\cal{L}}_{XY}=0.
\label{eq1001}
\end{equation}
 The pressure $p$ and the density $\rho$ are functions of $\psi$, $X$ and $Y$ through ${\cal{L}}$
\begin{equation}
p={\cal{L}} ,
\label{eq0042}
\end{equation}
\begin{equation}
\rho=2X {\cal L}_X+2Y{\cal L}_Y -{\cal L}.
\label{eq0043}
\end{equation}
The standard definition of the adiabatic speed of sound is
\begin{equation}
 c_s^2=\left.\frac{\partial p}{\partial\rho}\right|_{s/n} \, ,
\label{eq2601}
\end{equation}
where the differentiation is taken  at constant $s/n$,
i.e. for an isentropic process.
Here  $s=S/V$ is the entropy density and $n=N/V$  the particle number density
associated with the particle number $N$. 
We use the terminology and notation of Landau and Lifshitz \cite{landau}
(see also \cite{tsagas}).
Our Lagrangian is a function of three variables, namely 
$X$, $Y$ and $\psi$. Hence
\begin{equation}
 c_s^2=\left.\frac{\delta p}{\delta\rho}\right|_{s/n}
 =\left.\frac{(\partial p/\partial X)\delta X +(\partial p/\partial Y)\delta Y
+(\partial p/\partial \psi)\delta\psi}{(\partial \rho/\partial X)\delta X +(\partial \rho/\partial Y)\delta Y
+(\partial \rho/\partial \psi) \delta\psi}
 \right|_{s/n}   ,
\label{eq3601}
\end{equation}
where the differentials $\delta X$, $\delta Y$ and $\delta\psi$
are subject to the constraint $\delta(s/n)=0$.
Next we show that this constraint implies $\delta\psi=0$.

Our fluid is a two component system
with $p=p_1+p_2$, $\rho=\rho_1+\rho_2$.
As in \cite{bilic?} we start from the standard thermodynamic relation
for each component $i=1,2$
\begin{equation}
 \delta(\rho_i V)=
 T \delta S_i -p_i\,\delta V .
\label{eq1601}
\end{equation}
Assuming that there exist a conserved particle number  $N_i$ for each fluid,
the volume may be expressed in terms of particle number densities 
$ n_i=N_i/V$
 so 
\begin{equation}
 \delta V=-V\frac{\delta n_i}{n_i} .
 \label{eq0034}
\end{equation}
Equations (\ref{eq1601}) may then be written in the form
\begin{equation}
 \delta h_i= T \delta\left(\frac{s_i}{n_i}\right) +\frac{1}{n_i}\delta p_i  ,
\label{eq1603}
\end{equation}
where 
\begin{equation}
 h_i=\frac{p_i+\rho_i}{n_i}
\label{eq2605}
\end{equation}
is the enthalpy per particle. 
In the case of two conserved particle numbers, for an isentropic process 
we must have
$\delta (s/n_i)=0$ for both $i=1$ and $i=2$, where $s=s_1+s_2$.
As a consequence
\begin{equation}
 n_1 \delta\left(\frac{s_1}{n_1}\right) 
 +n_2 \delta\left(\frac{s_2}{n_2}\right)=
 -s_2 \frac{n_1}{n_2} \delta\left(\frac{n_2}{n_1}\right)=0 ,
\label{eq0035}
\end{equation}
where $\delta (n_2/n_1)$ vanishes because of (\ref{eq0034}).
Using this,  from (\ref{eq1603}) it follows
\begin{equation}
 \delta p|_{s/n}=n_1\delta h_1+n_2\delta h_2,
\label{eq0036}
\end{equation}
where $p=p_1+p_2$.

Furthermore, for an  isentropic relativistic flow
one can define the velocity potentials $\phi_i$ such that \cite{landau}
\begin{equation}
 h_iu_{i\mu}=\phi_{i,\mu} \, .
\label{eq2604}
\end{equation}
 Comparing this with (\ref{eq5021})
and identifying 
\begin{equation}
 \phi_1\equiv \Phi,
\hspace{1cm}
\phi_2\equiv \Theta ,
\label{eq1602}
\end{equation}
we find
\begin{equation}
 h_1=\sqrt{X}\,;
\hspace{1cm}
h_2=\sqrt{Y} .
\label{eq2602}
\end{equation}
Using this in (\ref{eq0036})
 we obtain
\begin{equation}
 \delta p|_{s/n}=\frac{n_1}{2\sqrt{X}}\delta X+\frac{n_1}{2\sqrt{Y}}\delta Y .
\label{eq0037}
\end{equation} 
Comparing this equation with
the general expression for the total differential of $p$
\begin{equation}
 \delta p=\frac{\partial p}{\partial X}\delta X +\frac{\partial p}{\partial Y}\delta Y
+\frac{\partial p}{\partial \psi}\delta\psi
\label{eq2607}
\end{equation}
%(\ref{eq3601})
 we conclude that an isentropic process implies
\begin{equation}
 \delta\psi=0
\label{eq0038}
\end{equation}
 or $\psi={\rm const}$.
Furthermore, from (\ref{eq0037}) it follows
\begin{equation}
 n_1=2\sqrt{X} {\cal L}_X ,
\label{eq2603}
\end{equation} 
\begin{equation}
 n_2=2\sqrt{Y} {\cal L}_Y .
\label{eq3603}
\end{equation} 
These two expressions  are
derived  assuming an isentropic process, i.e., keeping  $\psi={\rm const}$.
%Equation (\ref{eq0038}) implies
% \begin{equation}
% \delta\Theta+\tilde{\lambda}\Phi\delta\Phi
%\label{eq0039}
%\end{equation} 
If we had $\psi={\rm const}$ in (\ref{eq2009}), i.e., if  ${\cal L}={\cal L}(X,Y)$,
equations (\ref{eq2603})  and (\ref{eq3603}) would follow from
the field equation for $\Phi$ and $\Theta$
\begin{equation}
 ({\cal L}_X g^{\mu\nu}\Phi_{,\mu})_{;\nu}=0; \hspace{1cm}
 ({\cal L}_Y g^{\mu\nu}\Theta_{,\mu})_{;\nu}=0,
\label{eq2702}
\end{equation}
 which would imply  conservation of two currents %\cite{muk,bab}
\begin{equation}
 j_{1\mu}=2{\cal L}_X \Phi_{,\mu} =n_1u_\mu \, ; \hspace{1cm}
 j_{2\mu}=2{\cal L}_Y \Theta_{,\mu} =n_2u_\mu .
\label{eq2802}
\end{equation}
The particle number densities $n_1$ and $n_2$ in these  expressions coincide 
with (\ref{eq2603}) and (\ref{eq3603}).
However, in a more general case  ${\cal L}={\cal L}(X,Y,\psi)$,
the field equations for $\Phi$ and $\Theta$ do not imply 
conservation of the two
currents 
in (\ref{eq2802}).
Nevertheless,  equations (\ref{eq2603}) 
are still valid expressions for
conserved 
number densities when the condition $\delta(s/n)=0$ is imposed.

 The adiabatic speed of sound is now given by
\begin{equation}
 c_s^2=\left.\frac{\delta p}{\delta\rho}\right|_{\psi} 
 =\frac{{\cal L}_X(\delta X/\delta Y) +{\cal L}_Y}{({\cal L}_X +2 X {\cal L}_{XX})
(\delta X/\delta Y)
+ {\cal L}_Y +2Y{\cal L}_{YY}} 
 \label{eq3602}
\end{equation}
The ratio $(\delta X/\delta Y)$ may be expressed in terms of $X$, $Y$, and the derivatives of $\cal{L}$
using  the condition
\begin{equation}
\frac{\delta (\sqrt{X}{\cal L}_X)}{\sqrt{X}{\cal L}_X}=
\frac{\delta (\sqrt{Y}{\cal L}_Y)}{\sqrt{Y}{\cal L}_Y}
\label{eq3044}
\end{equation} 
which follows from (\ref{eq0034}), (\ref{eq2603}), and (\ref{eq3603}). We find
\begin{equation}
\frac{\delta X}{\delta Y} = \frac{X{\cal L}_X ({\cal L}_Y +2 Y {\cal L}_{YY})}{
 Y{\cal L}_Y({\cal L}_X +2Y{\cal L}_{XX})} 
 \label{eq4602}
\end{equation}
Using this we obtain
\begin{equation}
 c_s^2
 =\frac{X{\cal L}_X^2 ({\cal L}_Y +2Y{\cal L}_{YY}) +Y {\cal L}_Y^2({\cal L}_X +2 X {\cal L}_{XX})}{
(X{\cal L}_X+Y {\cal L}_Y)({\cal L}_X +2 X {\cal L}_{XX})
 ({\cal L}_Y +2Y{\cal L}_{YY})} 
 \label{eq5602}
\end{equation}
This expression may be written in the form
\begin{equation}
 c_s^2
 =c_{s1}^2\frac{\rho_1+p_1}{\rho+p} +c_{s2}^2\frac{\rho_2+p_2}{\rho+p} 
 %\label{eq6602}
\end{equation}
where $c_{s1}$ and $c_{s2}$ are defined as
\begin{equation}
c_{s1}^2=\frac{{\cal L}_X}{{\cal L}_X+2X{\cal L}_{XX}} ;
\hspace{1cm}
c_{s2}^2=\frac{{\cal L}_Y}{{\cal L}_Y+2Y{\cal L}_{YY}} .
%\label{eq014}
\end{equation}
\begin{equation}
 \rho_1+p_1=2X{\cal L}_X ;
 \hspace{1cm}
 \rho_2+p_2=2Y{\cal L}_Y.
%\label{eq6603}
\end{equation}


\begin{thebibliography}{99}
\bibitem{arkani}
N.~Arkani-Hamed, S.~Dimopoulos, and G.~Dvali,
Phys.\ Lett.\ B {\bf 429}, 263 (1998).
%\bibitem{turn1} Turner M S, Steigman G and Krauss L 1984
%{\it Phys.\ Rev.\ Lett.} {\bf 52} 2090
\bibitem{antoniadis} 
  I.~Antoniadis, N.~Arkani-Hamed, S.~Dimopoulos and G.~R.~Dvali,
  %``New dimensions at a millimeter to a Fermi and superstrings at a TeV,''
  Phys.\ Lett.\ B {\bf 436}, 257 (1998)
  [hep-ph/9804398].
%;
\bibitem{randall1}
L.~Randal and R.~Sundrum, Phys.\ Rev.\ Lett. {\bf 83}, 3370 (1999)
%Peebles P J E 1984 {\it Astrophys.\ J.} {\bf 284} 439
\bibitem{randall2}
L.~Randal and R.~Sundrum, Phys.\ Rev.\ Lett. {\bf 83}, 4690 (1999)
\bibitem{flanagan} 
  E.~E.~Flanagan, S.~H.~H.~Tye and I.~Wasserman,
  %``Cosmological expansion in the Randall-Sundrum brane world scenario,''
  Phys.\ Rev.\ D {\bf 62}, 044039 (2000)
  [hep-ph/9910498].
\bibitem{binetruy} 
  P.~Binetruy, C.~Deffayet and D.~Langlois,
  %``Nonconventional cosmology from a brane universe,''
  Nucl.\ Phys.\ B {\bf 565}, 269 (2000)
  [hep-th/9905012]. 
\bibitem{godlowski} 
  W.~Godlowski and M.~Szydlowski,
  %``Brane universes tested by supernovae Ia,''
  Gen.\ Rel.\ Grav.\  {\bf 36}, 767 (2004)
  [astro-ph/0404299].
\bibitem{kim1} 
J.E.~Kim, G.B.~Tupper, and R.D.~ Viollier, Phys.\ Lett.\ B {\bf 593}, 209 (2004);
\bibitem{kim2}
J.E.~Kim, G.B.~Tupper, and R.D.~Viollier, Phys. Lett. B {\bf 612}, 293 (2005).
\bibitem{bagger} 
  J.~Bagger and M.~Redi,
  %``Radion effective theory in the detuned Randall-Sundrum model,''
  JHEP {\bf 0404}, 031 (2004)
%  [hep-th/0312220].
\bibitem{goldberger} 
  W.~D.~Goldberger and M.~B.~Wise,
  %``Modulus stabilization with bulk fields,''
  Phys.\ Rev.\ Lett.\  {\bf 83}, 4922 (1999)
  [hep-ph/9907447].
\bibitem{csaki} 
  C.~Csaki, M.~L.~Graesser and G.~D.~Kribs,
  %``Radion dynamics and electroweak physics,''
  Phys.\ Rev.\ D {\bf 63}, 065002 (2001)
  [hep-th/0008151]. 
\bibitem{cheung} 
  K.~Cheung and T.~-C.~Yuan,
  %``Could the excess seen at 124-126 GeV be due to the Randall-Sundrum Radion?,''
  Phys.\ Rev.\ Lett.\  {\bf 108}, 141602 (2012)
  [arXiv:1112.4146 [hep-ph]].  
 \bibitem{bilic3} 
  N.~Bili\'c and G.~B.~Tupper,
  %``'Warm' Tachyon Matter from Back-reaction on the Brane,''
  arXiv:1302.0955 [hep-th]. 
\bibitem{silverstein}
  E.~Silverstein and D.~Tong,
  %``Scalar speed limits and cosmology: Acceleration from D-cceleration,''
  Phys.\ Rev.\ D {\bf 70}, 103505 (2004)
  [hep-th/0310221].
\bibitem{abramo}
  L.~R.~W.~Abramo and F.~Finelli,
  %``Cosmological dynamics of the tachyon with an inverse power-law potential,''
  Phys.\ Lett.\ B {\bf 575}, 165 (2003)
  [astro-ph/0307208].
\bibitem{felder} 
  G.~N.~Felder, L.~Kofman and A.~Starobinsky,
  %``Caustics in tachyon matter and other Born-Infeld scalars,''
  JHEP {\bf 0209}, 026 (2002)
  [hep-th/0208019].
\bibitem{kam9} A.~Kamenshchik, U.~Moschella, and V.~Pasquier, 
Phys.\ Lett.\ B {\bf 511}, 265 (2001).
\bibitem{bilic} N.~Bili\'c, G.B.~Tupper, and R.D.~Viollier,
Phys.\ Lett.\ B {\bf 535}, 17 (2002).
\bibitem{fab19} J.C.~Fabris,  S.V.B.~Gon\c calves, and P.E.~de Souza,
Gen.\ Relativ.\ Gravit.\ {\bf 34}, 53 (2002);
J.C.~Fabris, S.V.B.~Gon\c calves, and P.E.~de Souza,
Gen.\ Relativ.\ Gravit. {\bf 34}, 2111 (2002).
\bibitem{ben32}
M.C.~Bento, O.~Bertolami, and A.A.~Sen,
  %``Generalized Chaplygin gas, accelerated expansion and dark energy-matter
  %unification,''
Phys.\ Rev.\ D  {\bf 66}, 043507 (2002).
%  [arXiv:gr-qc/0202064].
\bibitem{mak11} M.~Makler, S.Q.~de Oliveira, and I.~Waga,
Phys. Lett. B {\bf 555}, 1 (2003);
 R.R.R.~Reis, M.~Makler, and I.~Waga,
Phys. Rev. D {\bf 69}, 101301 (2004).
  \bibitem{avelino} 
  A.~Avelino, N.~Cruz and U.~Nucamendi,
  %``Testing the EoS of dark matter with cosmological observations,''
  arXiv:1211.4633 [astro-ph.CO].
\bibitem{frenk} 
  C.~S.~Frenk and S.~D.~M.~White,
  %``Dark matter and cosmic structure,''
  Annalen Phys.\  {\bf 524}, 507 (2012)
  [arXiv:1210.0544].
\bibitem{primack} 
  J.~R.~Primack,
  %``Cosmology: small scale issues revisited,''
  New J.\ Phys.\  {\bf 11}, 105029 (2009)
  [arXiv:0909.2247]. 
\bibitem{boyanovsky} 
  D.~Boyanovsky and J.~Wu,
  %``Small scale aspects of warm dark matter : power spectra and acoustic oscillations,''
  Phys.\ Rev.\ D {\bf 83}, 043524 (2011)
  [arXiv:1008.0992].
\bibitem{viel} 
  M.~Viel, G.~D.~Becker, J.~S.~Bolton and M.~G.~Haehnelt,
  %``Warm Dark Matter as a solution to the small scale crisis: new constraints from high redshift Lyman-alpha forest data,''
  Physical Review D {\bf 88}, no. 4, 043502 (2013)
  [arXiv:1306.2314 [astro-ph.CO]].
\bibitem{schneider} 
  A.~Schneider, D.~Anderhalden, A.~Maccio and J.~Diemand,
  %``Warm Dark Matter: The End is Nigh,''
  arXiv:1309.5960 [astro-ph.CO]. 
  
\bibitem{coles}
P.~Coles and F.~Luchin, {\it Cosmology}, John Wiley \& Sons, Chichester (2002).
\bibitem{bilic?} 
  N.~Bili\'c G.~B.~Tupper and R.~D.~Viollier,
  %``Cosmological tachyon condensation,''
  Phys.\ Rev.\ D {\bf 80}, 023515 (2009)
  [arXiv:0809.0375 [gr-qc]].
\bibitem{landau} L.D. Landau, E.M. Lifshitz, {\it Fluid Mechanics}, Pergamon, Oxford
(1993).
\bibitem{tsagas} C.G. Tsagas, A. Challinor, and R. Maartens,
%``Relativistic cosmology and large-scale structure,''
Phys. Rep.  {\bf 465}, 61 (2008); [arXiv: 0705.4397].
\end{thebibliography}
\end{document}